\documentclass[prd,twocolumn,nofootinbib,superscriptaddress,10pt]{revtex4}

\usepackage{graphicx}
\usepackage{graphics}
\usepackage{amsmath,amssymb}
\usepackage[usenames]{color}
\usepackage{ulem}

\newcommand{\be}{\begin{equation}}
\newcommand{\ee}{\end{equation}}
\newcommand{\bea}{\begin{eqnarray}}
\newcommand{\eea}{\end{eqnarray}}
\newcommand{\beq}{\begin{equation}}
\newcommand{\eeq}{\end{equation}}
\newcommand{\beqa}{\begin{eqnarray}}
\newcommand{\eeqa}{\end{eqnarray}}

\newcommand{\pT}{{p_T}}

\newcommand{\antikT}{{anti-k$_T$}}
\newcommand{\Nvtx}{{N_{\rm vtx}}}
\definecolor{BrickRed}{cmyk}{0,0.89,0.94,0.28}%%%PANTONE 1805
\definecolor{MidnightBlue}{cmyk}{0.98,0.13,0,0.43}%%%PANTONE 302
\definecolor{DarkGreen}{rgb}{0.100806,0.495968,0.209979}
\definecolor{orange}{rgb}{0.587167,0.354498,0.146197}

\begin{document}
%\baselineskip=18pt
%\linespread{1.3}
%\begin{titlepage}
%\begin{center}
%%%%%%%%%%%%%%%%%%%%%%%%%%%%%%%%%%%%%%%%%%%%%%%%%%%%%%%%%%%%%%%%%%%%%%%%%%%%

\title{A data-driven method of pile-up correction for the substructure of massive jets}

\author{Raz Alon} \address{Department of Particle Physics \&
Astrophysics, Weizmann Institute of Science, Rehovot 76100, Israel}

\author{Ehud Duchovni } \address{Department of Particle Physics \&
Astrophysics, Weizmann Institute of Science, Rehovot 76100, Israel}

\author{Gilad Perez} \address{Department of Particle Physics \&
Astrophysics, Weizmann Institute of Science, Rehovot 76100, Israel}

\author{Aliaksandr P. Pranko}
\address{Lawrence Berkeley National Laboratory, Berkeley, CA 94720}

\author{Pekka K. Sinervo, F.R.S.C.} \address{Department of Physics, University of Toronto, 60 St. George Street, Toronto, Canada  M5S 1A7}

%\date{\today}

\begin{abstract}
We describe a method to measure and subtract the incoherent component of energy flow arising from multiple interactions from jet shape/substructure observables of ultra-massive jets. The amount subtracted is a function of the jet shape variable of interest and not a universal property. Such a correction is expected to significantly reduce any bias in the corresponding distributions generated by the presence of multiple interactions, and to improve measurement resolution. Since in our method the correction is obtained from the data, it is not subject to uncertainties coming from the use of theoretical calculations and/or Monte Carlo event generators. We derive our correction method for the jet mass, angularity and planar flow.
We find these corrections to be in good agreement with data on massive jets observed by the CDF collaboration. Finally, we comment on the linkage with the concept of jet area and jet mass area. 
\end{abstract}
%%\pacs{}

\maketitle

%%%%%%%%%%%%%%%%%%%%%%%%%%%%%%%%%%%%%%%%%%%%%%%%%%%%%%%%%%%%%%%%%%%%%%%%%%%%

\section{Introduction}
Incoherent processes in high-energy hadron-hadron collisions like multiple interactions, the underlying event in a high transverse momentum ($\pT$)\ scatter or instrumental effects may blur the picture when various hard processes are under study. This is especially  important for studies of high $\pT$\ ultra-massive jets:  
Though the jet substructure can be computed perturbatively with reasonable accuracy,
these incoherent processes lead to reductions in the resolution and sensitivity of various searches for new  physics~\cite{Dokshitzer:1998pt,Cacciari:2007fd,Cacciari:2009dp,Cacciari:2008gd,Rubin:2010fc,Salam:2009jx}. The existing correction methods rely mostly on the Monte Carlo simulation of the underlying event and additional (pile-up) interactions at high instantaneous luminosities.

We propose a data-driven method that enables one to measure the effect of incoherent contributions to jet substructure variables and to get an analytical expression for the functional form of the correction. Using this method one has available jet-variable dependent corrections rather than global ones. 
Thus, the measured substructure distribution should correspond to that arising from the hard part of the event. 
The correction technique
can be applied simultaneously to several jet-shape variables (of a fixed large mass) leading to improved resolution of the relevant jet variables and to an increase in the sensitivity to new physics signals. The proposed method has been successfully demonstrated with CDF data collected in proton-antiproton collisions at $\sqrt{s}=1.96$~TeV~\cite{CDFnew}.  

The susceptibility of modern jet algorithms that are  infra-red and collinear (IRC) safe to soft and weakly correlated contributions can be elegantly described by the concept of a jet area ~\cite{Cacciari:2007fd,Cacciari:2008gd,Salam:2009jx}. 
Such contributions may shift the value of any given jet variable. 
When studying the substructure of highly boosted ultra-massive jets of particular interest is the corrections to jet variables as a function of its value on a jet-by-jet basis, which is independent of the average global shift to its momenta.
The proposed method is based on a data-driven measurement of the size and effect of the incoherent component of energy flow for a given jet. 

 The actual measurement uses the method~\cite{Albrow:2006rt} of employing the dominant dijet topology of high $\pT$\ jets produced via QCD interactions and measuring the energy deposition in a fixed size cone rotated by $90^o$ relative to the dijet axis (as is used in a recent CDF study~\cite{ourCDF,CDFnew}).  
Our technique can be applied to both high and low instantaneous luminosity ($L$)\ regimes such as those experienced or expected at the Tevatron and the Large Hadron Collider.
In the examples we use to illustrate this procedure, the average dependence of the corrections on the relevant variables has been determined by the CDF collaboration using a large sample of high $\pT$\ jets.  
This technique gives the actual correction to be applied to the relevant jet variables.  
However, in practice, the corrections are parametrized based on theoretical expectations, as we discuss below.

The size of the incoherent effects can be also extracted using sophisticated methods that have been studied in~\cite{Cacciari:2007fd,Cacciari:2008gd,Salam:2009jx}\ and that  are incorporated within the FastJet framework~\cite{fastjet, FastJetWeb}.
In the latter case, assuming a diffuse soft component~\cite{Cacciari:2007fd}, one can determine a  correction by measuring the energy density of the soft component in the event, multiplying it by the active jet area and then estimating the corresponding shift in the value of the jet shape variable under study. The case of passive area proceeds in a similar manner and is further discussed below in the context of jet mass area~\cite{Sapeta:2010uk}. 

In the following, we first describe the general procedure, outline the expected corrections for mass, angularity and planar flow, and illustrate how the CDF data confirm our method predictions.  
We then comment on the relationship of our technique to the concept of jet mass area.

\section{The general prescription}
Consider a jet-shape variable $X$ that characterizes the energy flow within ultra-massive highly boosted jets whose transverse momenta and invariant mass are in a given predefined range. Below we focus on the high jet mass region ($> 70$~GeV) since the QCD contribution is better controlled there, and since such massive jets are of special importance for  various new physics searches. 

We evaluate the variation of $X$\ under the additional incoherent component of radiation
\beqa
\hspace*{-.1cm}\Delta X\big|_{p_J,m_J} = { \partial X\over \partial m_J}\big|_{p_J,m_J}\delta m_J +  \sum_{i\in R^{90^o}} { \partial X\over \partial E_i}\big|_{p_J,m_J}\delta E_i \,,\label{Master}
\eeqa
where $p_J$ is the jet momenta (or transverse momenta for hadronic collider)\ and the summation $\sum_{i\in R}$ corresponds to the sum of the energy of calorimeter cells  ($E_i$)  inside a jet with a size-parameter $R$.  The summation $\sum_{i\in R^{90^o}}$ corresponds to the sum of energy deposited in a cone of area $a_0=\pi R^2$ whose axis is rotated by $90^o$ in $\phi$ direction.
It is assumed here that $X$ is measured in the leading jet and that the incoherent energy deposition inside the leading jet is equal to that observed, at least on average, to the cone perpendicular in azimuth: $\sum_{i\in R}$=$\sum_{i\in R^{90^o}}$.
It is worth mentioning here again that the method is independent of the way the additional incoherent component of energy is measured.   This procedure will work for any IRC jet algorithm and as long as $R^2\ll1$\ (as we work to leading order).

Generally, the correction to $X$ ($\Delta X$) can be written as a function of $X$ itself   
for the variables we are interested in, so that
\beqa
\Delta X(p_J,m_J)=f(X,p_J,m_J) \delta m_J^2\oplus g(X,p_J,m_J) \delta E\,,
\eeqa
where $f(X,p_J,m_J)$ and  $g(X,p_J,m_J)$ are analytic functions that  are computed below for few jet-variables, and the multiplicative coefficients for $\delta m_J^2$ and $\delta E$ can be determined from the data.

The correction procedure for jet mass, angularity and planar flow are derived below. 
The procedure gives rise to concrete predictions of the form of the corrections ($\Delta X(X,p_J,m)$)  as a function of the value of the jet-variable.  
Because the corrections can be determined directly from the data, their uncertainties are relatively small and can be controlled experimentally.

\section{Subtraction method for jet mass}
This case is a simplification of the general case described by Eq. \eqref{Master}, since
$X$ is one of the two variables we normally control independently.
Nevertheless, in order to demonstrate the procedure we analyze it in some length. 
The correction to the jet mass is:
\beqa
\Delta m_J\big|_{p_J,m_J} =  \sum_{i\in R^{90^o}} { \partial m\over \partial E_i}\big|_{p_T,m_J}\delta E_i \,.
\eeqa
To estimate the right-hand-side (RHS) of this relation note that the jet mass squared is given by
$m_J^2= \left(\sum_{i\in R}P_i\right)^2$,
and so the correction to it is
\beq\label{Delm}
\Delta m_J^2 \sim p_J \sum_{i\in R^{90^o}}\delta E_i \theta_i^2\equiv  \sum_{i\in R^{90^o}} \delta m^2_i\,.
\eeq
Since to leading order $\Delta m_J^2=2 m_J \delta m_J$
we find that the leading order correction to the jet mass is given by (for a related discussion see~\cite{Salam:2009jx}) 
\beq\label{delmJ}
\delta m_J \sim   \sum_{i\in R^{90^o}} { \delta m^2_i \over 2 m_J}   \,.
\eeq
We thus find that for a fixed $p_T$ the correction to the jet mass is proportional to the inverse of that mass and the coefficient can  
be fit from the data. This is in agreement with the CDF results for the Midpoint, \antikT\ \cite{Cacciari:2008gp}\ or Midpoint/SC (Midpoint using search cones) jet algorithms~\cite{CDFnew}. 
In this case, data were analyzed separately for events with one primary vertex  ($\Nvtx = 1$) and for events with multiple interactions ($\Nvtx >1$) (i.e. single and  multiple interactions events).
The  $\Nvtx>1$\ corrections behave as expected from the analysis above.  
Furthermore, the $\Nvtx=1$\ corrections show the average effect of the underlying event in the hard scatter on the jet mass, but may not accurately represent the true behaviour of the soft component given that our calculation assumes it behaves incoherently.  
The difference between the two corrections separates out the purely incoherent component, and gives further confirmation that the multiple interactions act purely incoherently, scaling with both the level of multiple interactions and having the appropriate $R^4$\ dependence on the jet radius.

\begin{figure}
\begin{center}
\includegraphics[width=7.7cm]{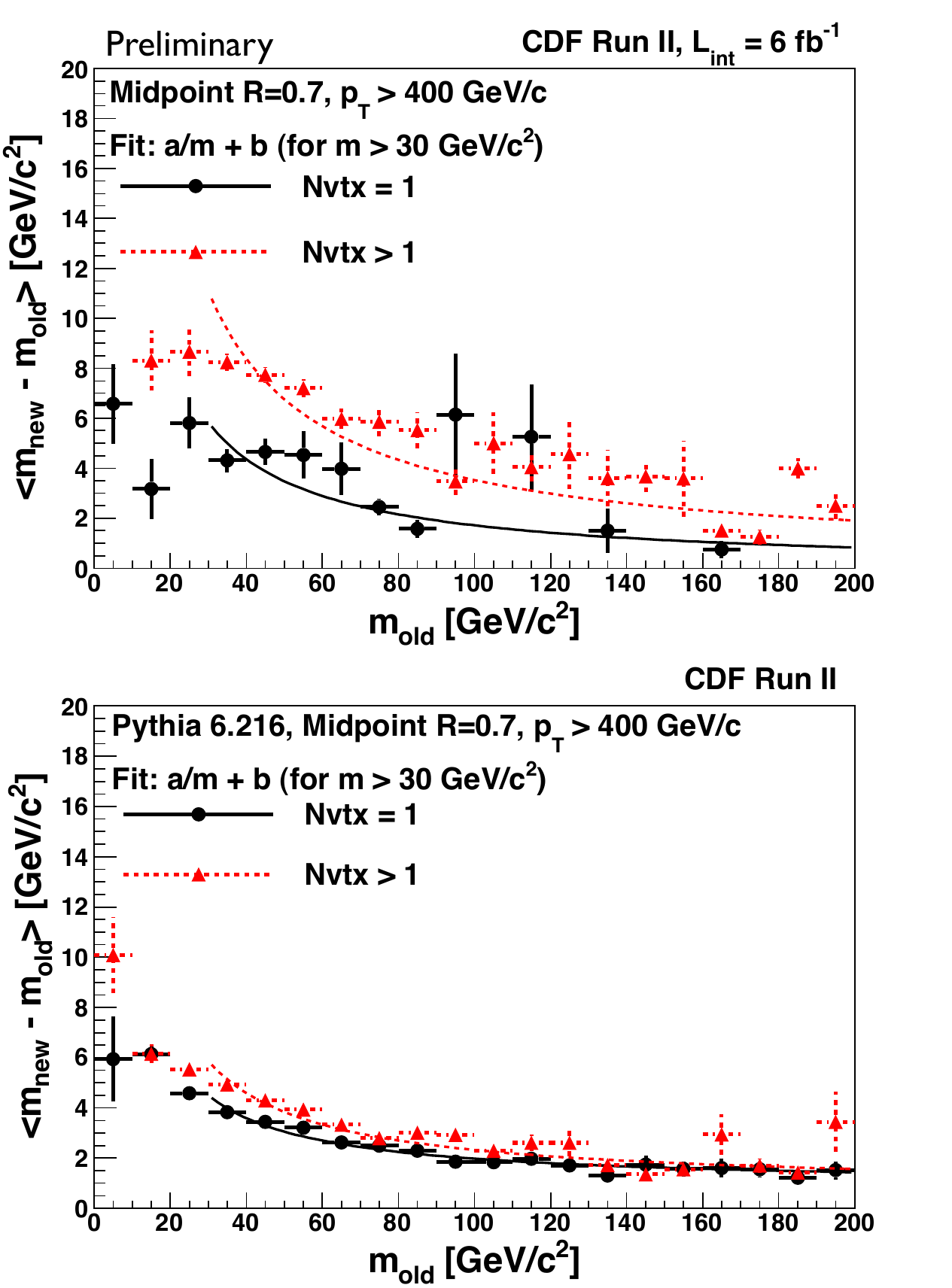}
\caption{On the upper panel we show the CDF data and a fit based on the relation derived in Eq.~\eqref{delmJ}. The data collected had on average $\sim3$\ multiple interactions per event (including the hard interaction). On the lower panel we show the corresponding MC predictions including full detector simulation~\cite{CDFnew}. }
\label{fig:MMI}
\end{center}
\end{figure}

This is shown in Fig.~\ref{fig:MMI}\ from~\cite{CDFnew} which includes both the PYTHIA 6.216 Monte Carlo (MC) prediction \cite{PYTHIA}\ (including full detector simulation) and the fit to the functional dependence given in~Eq.~\eqref{delmJ}. 
The vertical axis corresponds to the average change in the jet mass upon adding the contributions from the 90$^o$ cone as a function of the measured jet mass (the horizontal axis). We do not expect the MC to provide a precise determination of the overall scale of the change but rather give insight towards the shape of the correction, since the statistics is much less of an issue in this case.
In particular, the $\Nvtx>1$\ contribution will be much less given that this MC calculation assumed only $\sim0.5$\  interactions in addition to the hard scatter per event.
The reader may note that the plot also includes the low mass region which is beyond the focus of the present study. 

%%%%%%%%%%%%%%%%%%%%%%%%%%%%%%%%%%%%%%%%%%%%%%%%%%%%%%%%%%%%%%%%%%%%%%%%%%%%
\section{Subtraction method for angularity}
The small angle expression for angularity is~\cite{Berger:2003iw,Almeida:2008yp}
\begin{equation}
\tau_a(R,p_T) \sim \frac{2^{a-1}}{m_J} \,\sum_{i \in jet} E_i\,\theta_i^{2-a} \,,
\label{tauadef}
\end{equation}
where $a\leq2$ is required for IRC safety. Recently, the $a=-2$ distribution was measured by CDF for jets with $p_T>400$~GeV
and mass in the window $90\leq m_J\leq 120$\,GeV, hence we will focus on this specific value of angularity (the procedure below should work, in principle, for arbitrary value of $a$, however, clearly $a=0$ is special since it is not independent of the jet mass variable.).
To leading order, the correction from incoherent energy deposition is given by
\beqa
\Delta \tau_a&=& { \partial \tau_a\over \partial m_J}\delta m_J +  \sum_{i\in R^{90^o}} { \partial \tau_a\over \partial E_i}\delta E_i \nonumber \\
&\simeq&
-{\tau^J_a\over 2m_J^2}  \sum_{i\in R^{90^o}} \delta m^2_i\ + \frac{2^{a-1}}{p_J m_J} \,\sum_{i\in R^{90^o}} \delta m^2_i\,\theta_i^{-a}
\nonumber \\
&=&
 \sum_{i\in R^{90^o}}{ \delta m^2_i\over 2m_J^2}  \left( \frac{2^{a}m_J}{p_J } \theta_i^{-a}\oplus\tau^J_a\right)\,,\label{AngMI}
\eeqa
where  we use Eq.~\eqref{Delm} to simplify the RHS.  We note that $\tau^J_a$ corresponds to the jet angularity before the correction. We also note that the two types of contributions should be added {\sout incoherently}\ in quadrature as indicated by the  $\oplus$ symbol. Eq.~\eqref{AngMI} implies that for a fixed jet mass (as is often applied in new physics searches) the leading order correction to the angularity consists of two terms: a constant and a term proportional to the value of the angularity itself.

Let us denote by $R_{12}$ the ratio between the second and the first terms in the parenthesis of the RHS of Eq. \eqref{AngMI}, 
\beq
R_{12}={2^a\over\tau_a^J} \frac{m_J}{p_J} { \sum_{i\in R^{90^o}} \delta m^2_i\,\theta_i^{-a} \over \sum_{i\in R^{90^o}} \delta m^2_i}  \,.
\eeq
The above ratio can be estimated by taking the minimum and maximum value for the angularity, $\left(\tau^J_a\right)^{\rm min,\,max}$, which may be obtained from the leading order perturbative QCD result~\cite{ourCDF,Almeida:2010pa},
\beq 
\left(\tau^J_a\right)^{\rm min}\simeq  \left(\frac{m_J}{2p_J}\right)^{1-a},\, \ \ \left(\tau^J_a\right)^{\rm max}\simeq 2^{a-1}\,R^{-a} \,{m_J\over p_J }.
\eeq
We therefore find that the ratio between the minimum and maximum contributions $\left(R_{12}^{\rm min,\,max}\right)_i$ is:
\bea
\left(R_{12}^{\rm min}\right)_i&\sim& 2\,\theta_i^{-a} \left(m_J\over p_J\right)^a  \sim 
2 \left(m_J\over R p_J\right)^a\,,\nonumber\\ 
\left(R_{12}^{\rm max}\right)_i&\sim& 2\,\theta_i^{-a} R^{a}  \sim 
2
 \,,\nonumber
\eea
where on the RHS we have used the approximation $\theta_i\sim R$ for the most important contributions. 

\begin{figure}
\begin{center}
\includegraphics[width=7.7cm]{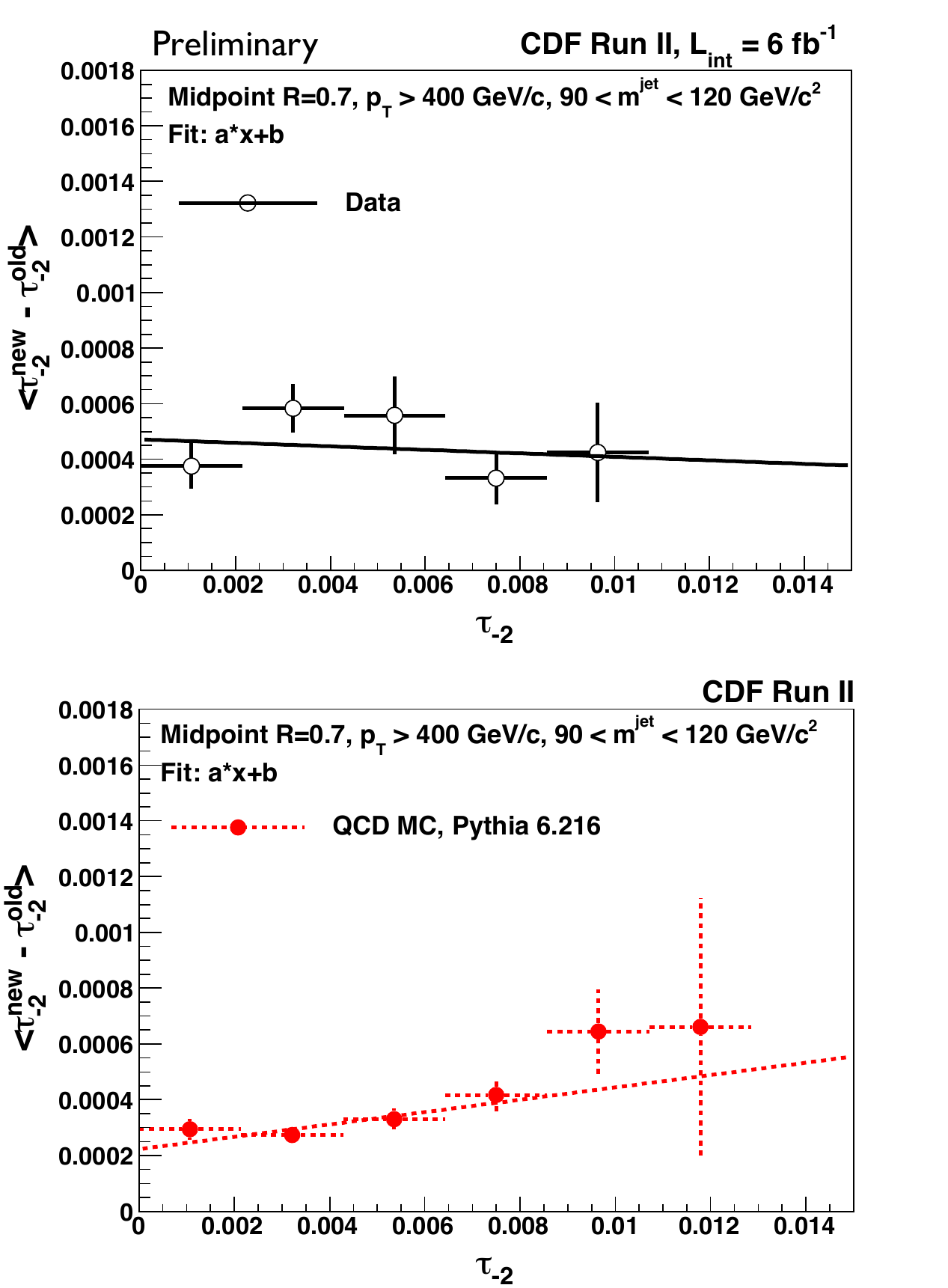}
\caption{On the upper panel we show the CDF data and a fit based on the relation derived in Eq.~\eqref{AngMI}. On the lower panel we show the corresponding MC predictions including full detector simulation~\cite{CDFnew}.}
\label{fig:Ang}
\end{center}
\end{figure}

The interesting angularity distributions, relevant to highly boosted massive jets, are those with negative $a$~\cite{ourCDF,Almeida:2008yp} which emphasize the radiation towards the edge of the cone.
Consequently, we find that over the interesting range of parameters the constant corection term dominates with some subdominant linear contribution towards $ \left(\tau^J_a\right)^{\rm max}$.
We also find that in general the relative correction to angularity is small
\beq
{\Delta \tau_a\over \tau_a}\sim  \sum_{i\in R^{90^o}}{ \delta m^2_i\over 2m_J^2} \left(R_{12}\right)_i \lesssim  
\sum_{i\in R^{90^o}}{ \delta m^2_i\over m_J^2}\sim {2\delta m_J\over m_J} \ll1.
\eeq
Analysis of the expected corrections at CDF shows that for $p_T\geq 400$\,GeV, $R=0.7$ and $m_J\sim 100$\,GeV then 
${\delta \tau_a\over \tau_a}\lesssim {2\times 4 \,\rm GeV/ 100\,GeV}={\cal O}(8\%)$, which is in a good agreement with the data~\cite{ourCDF}.
The measured correction, the PYTHIA 6.216 Monte Carlo (MC) prediction (including full detector simulation) and the fit to the functional dependence given by~Eq.~\eqref{AngMI} are shown in Fig.~\ref{fig:Ang}\ \cite{CDFnew}.
The vertical axis corresponds to the change in the angularity upon adding the contributions from the 90$^o$ cone as a function of the measured angularity (the horizontal axis). The small number of events after having imposed the high mass requirement does not allow us to separate out contributions from single interaction events and events with multiple interactions, as the data is dominated by $\Nvtx>1$\ events.
The form of the distribution is consistent with the prediction.

%%%%%%%%%%%%%%%%%%%%%%%%%%%%%%%%%%%%%%%%%%%%%%%%%%%%%%%%%%%%%%%%%%%%%%%%%%%
\section{Subtraction method for planar flow}
To define the planar flow, $Pf$~\cite{Almeida:2008tp,Almeida:2008yp,Thaler:2008ju}, we first construct, for a given jet, a $2\times2$ matrix  
$I_E$ 
\begin{equation}
I^{kl}_{E}=\frac {1}{m_J} \sum_{i\in R} E_i \frac{p_{i,k}}{E_i}\,\frac{p_{i,l}}{E_i}\, ,
\end{equation}
where $p_{i,k}$ is the $k^{th}$ component of the $i^{th}$ particle's  transverse momentum relative to the jet momentum axis.
We point out that at small angles $I_w$\ corresponds to a straightforward generalization of $\tau_0$, but promoted to a two-dimensional tensor 
\beq\label{tauxy}
\tau_0^{xy}\equiv \frac{1}{2m_J} \,\sum_{i \in jet} E_i\,\theta_i^x\theta^y={I_w\over 2}\,.
\eeq
We shall return to this point.
Given $I_{w}$, we define $Pf$ for that jet as
\begin{equation}
Pf = 4\, {\rm \frac{det(I_E)}{{\rm tr}(I_E)^2}} =
\frac{4 \lambda_1 \lambda_2}{(\lambda_1 + \lambda_2)^2} ,
\end{equation}
where $\lambda_{1,2}$ are the eigenvalues of $I_E$.

$I_E$ is a real symmetric matrix, so without loss of generality it can be expanded as a sum of three basis matrices
\beq
I_E= p_0 \,\sigma_0 + p_x\,\sigma_x +p_z\,\sigma_z\,,
\eeq
where $\sigma_0\equiv{\mathbf{1}}_2/\sqrt{2}$ ($\mathbf{1}_2$ is a unit matrix), $\sigma_{x,z}$ are the corresponding Pauli matrices and we use the normalization
${\rm tr}\left(\sigma_i \sigma_j\right)=\delta_{ij}$ such that the $\sigma_{i}$s form an orthonormal basis; finally, the $p_i$s are real numbers and the usefulness of the analogy with a two+one dimensional Lorentz group become clear since $Pf$ is now given by
\beq
Pf={p_0^2-p_i^2\over p_0^2}\equiv {m_{I_E}^2\over p_0^2}
\equiv {1\over \gamma_{I_E}^2}\equiv 1-\beta_{I_E}^2
\eeq
 with $p_i^2\equiv p_x^2+p_z^2$.
Let us first consider the contribution to $Pf$ from a single calorimeter cell. It satisfies the "null energy" condition of a massless particle $(p^1_0)^2-(p^1_i)^2=0$ where this is independent of the chosen frame in which $I_w$ is calculated. Note that this is the first point where our result deviates from a generic trivial description of symmetric real matrices.  
Thus $Pf$ actually corresponds to one over the boost factor for a system consisting of a set of massless particles in three dimensions, or to
the ratio of the invariant mass of a set of "massless particles" to their square of sum of energies. 

\begin{figure}[htbp]
\begin{center}
\includegraphics[width=7.7cm]{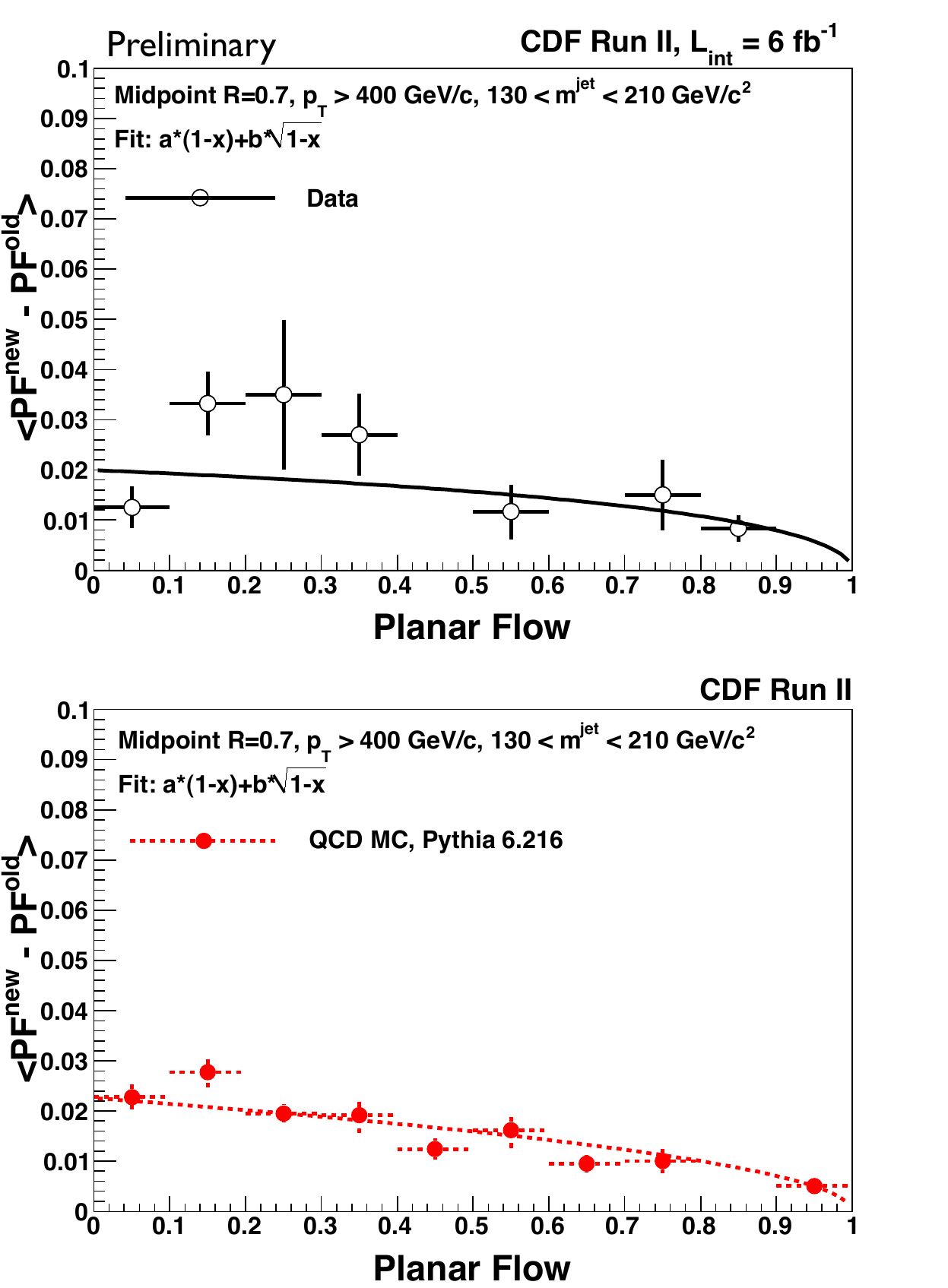}
\caption{On the upper panel we show the CDF data and a fit based on the relation derived in Eq.~\eqref{PFMI}.  On the lower panel we show the corresponding MC predictions including full detector simulation~\cite{CDFnew}. }
\label{fig:PF}
\end{center}
\end{figure}

Let us find the leading order correction due to incoherent energy depositions
\beqa 
\Delta Pf&=& {\partial  Pf\over \partial  p_0 }\delta p_0 +  {\partial  Pf\over \partial p_i }\delta p_i ={2\over p_0}\left(\beta_{I_E}^2\delta p_0-\beta_{I_E}\delta p_i\right)\nonumber \\ &=&{2\over p_0}\left[(1-Pf)\delta p_0-\sqrt{1-Pf}\,\delta p_i\right]
\eeqa
In order to obtain the value of $p_0$ in terms of observables we use Eq.~\eqref{tauxy}
\beq
p_0=\sqrt2\,\tau_0\,.
\eeq
While $\tau_0$ is a simple function of the jet mass and momenta, as explicitly obtained when evaluating the jet mass from its four momenta (assuming $m_J\ll P_J$ and $R\ll1$)
\beqa
m_J^2&\simeq& \left(P_J+\sum_{i\in R} {\delta p^2_i\over 2E_i}, P_J,\vec 0\right)^2 \approx P_J\sum_i {\delta p^2_i\over 2E_i}\nonumber \\ 
&\approx&
P_J\sum_i E_i \theta_i^2 = 2P_J m_J\,\tau_0\, \Rightarrow\, p_0\simeq{m_J\over \sqrt2\,P_J}\,.\label{p0mJ} 
\eeqa
We thus obtain the final and simple result for the planar flow correction,
\beqa 
\Delta Pf&=& {\sqrt2\, P_J\over m_J}\left[(1-Pf)\delta p_0\oplus\sqrt{1-Pf}\,\delta p_i\right]\,.\label{PFMI}
\eeqa
Let us estimate what is the expected size of $\delta p_{0,i}$. Since the correction from the incoherent radiation is random we generally expect 
$\delta p_i \sim \delta p_0$.  Using Eq.~\eqref{delmJ} and~\eqref{p0mJ}. we find 
\beq
\delta p_0 \simeq{\delta m_J\over \sqrt2\,P_J}\,.
\eeq
The largest correction is expected for $Pf\sim0$ which is roughly given by 
\beqa 
\Delta Pf^{\rm max}\sim {\sqrt2\, P_J\over m_J}\sqrt{\delta p_0^2+\delta p_0^2}\sim \sqrt2 \, {\delta m_J\over m_J}\,
\eeqa
For the CDF data we find $\Delta Pf\lesssim 7\%$\ for  $m_J\sim100\,$GeV.
The measured correction, the MC prediction (including full detector simulation) and the fit to the functional dependence given in~Eq.~\eqref{PFMI} is shown in Fig.~\ref{fig:PF} taken from the CDF data~\cite{CDFnew}.
The vertical axis corresponds to the change in the observed planar flow  as a function of the planar flow. As in the angularity case, contributions from single-vertex events are not separated given their small number.
The shape and normalization of the distribution is consistent with the prediction.

%%%%%%%%%%%%%%%%%%%%%%%%%%%%%%%%%%%%%%%%%%%%%%%%%%%%%%%%%%%%%%%%%%%%%%%%%%%
\section{Relation with jet areas}
Recently, the concept of ``jet area'' was introduced~\cite{Cacciari:2007fd} as a way of understanding the behaviour of jet observables in high instantaneous luminosity environments.
It was shown that once the jet's size becomes dynamical, as with modern IRC safe jet algorithms, this concept  turns out to be useful when assessing the susceptibility to incoherent energy contributions of various jet-variable measurements.
Our emphasize here is slightly different, as in~\cite{Cacciari:2007fd} we focus on applying data-driven corrections to jet-variable distributions over a large range of instantaneous luminosities, but we are also interested to get a semi-analytical understanding of the possible shape and size of the correction for each of the substructure variables.  
It is interesting to briefly mention the correspondence with the jet area concept, in particular in the context of the recent study of the ``mass area''~\cite{Sapeta:2010uk}.  
Our aim is two-fold: First, we show that knowing the jet mass and other shape variables such as angularity allows one to more precisely determine the jet mass area (and possibly other jet shape areas). Second, we argue that in the region of interest, the difference between jet mass area and jet area (of massless QCD events) is small, which implies that our method can be easily adapted by using a global extraction of the median energy density from data. 

We demonstrate our points explicitly using studies of the Midpoint and \antikT\  jet algorithms performed by the CDF collaboration~\cite{CDFnew}\ (the Midpoint results are essentially identical to those obtained with the SISCone algorithm~\cite{Salam:2007xv}, as expected since the two algorithms use similar split and merge procedure).
However, it is trivial to see that the same conclusions also applied to other jet algorithms. In the following we consider the ``passive jet area'' concept, where analytic results can be obtained. We focus on the the region with high degree of collimation for ultra massive jets, defined as $\epsilon^2\ll1$  where $\epsilon\equiv m/(p_T R)$. 
It is assumed that the boosted jet  consists of two partonic decay products of a heavy particle of mass $m$, which are well contained inside the jet (as is now qualitatively established by the CDF study of angularity~\cite{CDFnew}, this assumption holds also for QCD massive jets, basically doing the measurement for a fixed $\epsilon$). It is useful to define $a^0_j\equiv\pi R^2$ as the naive jet area of radius $R$, and following the definitions of~\cite{Sapeta:2010uk}  we define $\Delta_{12}$ as the rapidity-azimuth difference between two daughter particles, $x=\Delta_{12}/R$ and $z={\rm min}(p_{T^1},p_{T^2})/p_T$ in order to characterize the primary daughter particles. 

We find the following relation between $x$ and $z$ (assuming $R^2\ll1$)
\begin{equation}\label{cor}
x^2= {\epsilon^2\over z(1-z)}\,,
\end{equation}
and for later usage denote $z_1(x=1)\equiv \epsilon^2\big(1+\epsilon^2\big)+{\cal O}\big(\epsilon^6\big)\,.$
Let us begin with discussing the SISCone jet finder.  In this case one can minimize the area of the boosted jet by requiring the two daughter partons to be contained in a single jet. This is satisfied provided that $1<x<x_c\equiv 1/(1-z)$~\cite{Sapeta:2010uk}. 
The left inequality implies that $0<z<z_1$. On the other hand, maximizing the boosted jet area is achieved when the jet is the union of the three cones (around the mother and two daughter particles). 
This implies that  $x_c>x$, namely, $z>\epsilon^2\big(1-\epsilon^2\big)+{\cal O}\big(\epsilon^6\big)$. 
Since $\epsilon$\ is by construction very small, $z$\ cannot be changed significantly and leads, as anticipated, to small differences between the jet mass area and that of low mass jets analyzed in~\cite{Cacciari:2007fd}.
For the \antikT\ algorithm,  two interesting cases are found:  (1) for $1/(1+z)<x<1$\  the jet area is bigger then $a^0_j$, and
solutions are found for $8\epsilon^2<1$, implying that $z_1<z<\epsilon^2\big(1+3\epsilon^2\big)+{\cal O}\big(\epsilon^6\big)$.
(2) for $1<x<x_c$  the jet area is smallerer then $a^0_j$, and this coincides with the case analyzed above for SISCone again implying that for massive boosted collimated jets the expected deviation from the low mass jet area is small.  Thus, when studying high mass collimated jets one can approximately use either the $90^o$ cone method or derive the correction from the product of the jet area and the energy density following the prescription of~\cite{Cacciari:2007fd}.

Finally, we would like to mention that under the two-body approximation~\cite{Almeida:2008yp}, for a fixed momentum and mass the two-body jet's energy flow is fully characterized by a single continuous parameter. The asymmetry parameter $z$ is a simple function of the angularity or the soft particle distance from the jet axis~\cite{Almeida:2010pa} . 
It implies that given a jet mass and angularity one can extract the parameter $x$ in Eq.~\eqref{cor}), fully determining the jet mass area as defined in~\cite{Sapeta:2010uk}.

For example expanding the angularity variable away from the symmetric configuration ($z\ll1/2$) one finds
\beq 
z\sim \left({R\,\epsilon\over 2} \right)^3 \, \big(\tau_{-2}\big)^{-1}={\tau_{-2}^{\rm min}\over \tau_{-2}}\,,
\eeq
which, as expected, shows that as the angularity (which is supported by radiation towards the cone edge) increases so does the asymmetry ($z\to0$). 
The distribution of boosted massive jets originated both from QCD and massive particles (with 2-body decay) peak around the symmetric configuration~\cite{Almeida:2008yp}  (around $z=1/2$, unlike what is sometimes mentioned in the literature).  Therefore, it is useful to show the relation between $z$ and angularity in this region as well
\beq
z\sim {1\over2}\left(1-{1\over2}\sqrt{{\tau_{-2}\over \tau_{-2}^{\rm min}}-1}\,\right)\,,
\eeq
where, as expected we see that as the angularity departs from its minimal value the asymmetry parameter decreases from its maximum value accordingly. We emphasize again that the recent CDF study indeed qualitatively confirms the two body descriptions of massive jets and the peak around 
$\tau_{-2}^{\rm min}$ and a drop for larger values is clearly observed~\cite{CDFnew}.
It implies that one can further sharpen the extraction of the jet area via a measurement of its angularity.

\section{Conclusions}
To conclude we have provided a formalism to determine and take into account the effects due to incoherent (and approximately incoherent) radiation to various jet-variable distributions. 
We showed that the incoherent radiation induces jet-shape-dependent corrections, and showed how they can be calculated from collision data for jet mass, angularity and planar flow. We provided an analytic form for the corrections to these variables. 
These predictions have been supported by MC studies and have been verified by results from the  CDF collaboration~\cite{CDFnew}. Finally we also commented on the relation of our method with the concept of jet area.

\section*{Acknowledgments} 
We thank Gavin Salam and Sebastian Sapeta for useful discussions and comments on the manuscript. 
This work is supported in part by the Shrum Foundation at the Weizmann Instiutute of Science, the U.S. Department of Energy and the Canadian Natural Sciences and Engineering Research Council. 
We also thank the staff of the Ernest Orlando Lawrence Berkeley and Fermilab National Laboratories, where a part of this work was performed. 
GP is the Shlomo and Michla Tomarin career development chair and supported by the Israel
Science Foundation (grant \#1087/09), EU-FP7 Marie Curie, IRG
fellowship, Minerva and G.I.F., the GermanÐIsraeli Foundations, and the Peter \& Patricia Gruber Award.
%%%%%%%%%%%%%%%%%%%%%%%%%%%%%%%%%%%%%%%%%%%%%%%%%%%%%%%%%%%%%%%%%%%%%%%%%%%%

\end{document}